\documentstyle[pra,multicol,aps,amstex,epsfig]{revtex}
\begin{document}

\setlength{\textwidth}{150mm}
\setlength{\textheight}{240mm}
\setlength{\parskip}{2mm}

\input{epsf.tex}
\epsfverbosetrue

\draft

\renewcommand{\baselinestretch}{1.0}

\title{Quasiperiodic Envelope Solitons}

\author{Carl Balslev Clausen$^1$, Yuri S. Kivshar$^2$,
Ole Bang$^1$, and Peter L. Christiansen$^1$}

\address{
$^1$ Department of Mathematical Modelling,
Technical University of Denmark, DK-2800 Lyngby, Denmark\\
$^2$ Australian Photonics Cooperative Research Centre,
Research School of Physical Sciences and Engineering\\
Optical Sciences Centre, Australian National University,
Canberra ACT 0200, Australia}

\maketitle

\normalsize

\begin{abstract}
We analyse nonlinear wave propagation and cascaded self-focusing 
due to second-harmonic generation in {\em Fibbonacci optical 
superlattices} and introduce a novel concept of nonlinear physics, 
the {\em quasiperiodic soliton}, which describes spatially localized 
self-trapping of a quasiperiodic wave. 
We point out a link  between the quasiperiodic soliton and 
partially incoherent spatial solitary waves recently generated 
experimentally.
\end{abstract}


\begin{multicols}{2}
\narrowtext

For many years, solitary waves (or {\em solitons}) have been 
considered as {\em coherent localized modes} of nonlinear systems,
with particle-like dynamics quite dissimilar to the irregular and
stochastic behaviour observed for chaotic systems \cite{book}. 
However, about 20 years ago Akira Hasegawa, while developing a 
statistical description of the dynamics of an ensemble of plane 
waves in nonlinear strongly dispersive plasmas, suggested the 
concept of an {\em incoherent temporal soliton}, a localised 
envelope of random phase waves \cite{hasegawa}. 
Because of the relatively high powers required for generating 
self-localised random waves, this notion remained a theoretical 
curiosity until recently, when the possibility to 
generate spatial optical solitons by a partially incoherent source 
was discovered in a photorefractive medium \cite{inc_exp}, known to 
exhibit strong nonlinear effects at low powers.

The concept of incoherent solitons can be compared with a different 
problem: the propagation of a soliton through a spatially 
disordered medium. 
Indeed, due to random scattering on defects, the phases of the 
individual components forming a soliton experience random 
fluctuations, and the soliton itself becomes {\em partially 
incoherent} in space and time.
For a low-amplitude wave (linear regime) spatial incoherence is
known to lead to a fast decay. 
As a result, the transmission coefficient vanishes exponentially 
with the length of the system, the phenomenon known as Anderson 
localisation \cite{gredeskul}. 
However, for large amplitudes (nonlinear regime), when the 
nonlinearity length is much smaller than the Anderson localization 
length, a soliton can propagate almost unchanged through a 
disordered medium as predicted theoretically in 1990 
\cite{my_prl} and recently verified experimentally \cite{hopkins}.

These two important physical concepts, spatial self-trapping of light 
generated by an incoherent source in a homogeneous medium, and 
suppression of Anderson localisation for large-amplitude waves in 
spatially disordered media, both result from the effect of 
strong nonlinearity.
When the nonlinearity is sufficiently strong it acts as {\em an 
effective
phase-locking mechanism} by producing a large frequency shift of the
different random-phase components, and thereby introducing {\em an 
effective order} into an incoherent wave packet, thus enabling the 
formation of localised structures. In other words, both phenomena 
correspond to the limit when the ratio of 
the nonlinearity length to the characteristic length of (spatial or 
temporal) fluctuations is small. 
In the opposite limit when this ratio is large the wave 
propagation is basically linear.

{\em What will happen in the intermediate case when the length scales 
of nonlinearity and fluctuations become comparable ?} 
It is usually believed that localised structures would not be able to 
survive for such incoherent wave propagation and should rapidly decay.
In this Letter we show that, at least for aperiodic inhomogeneous 
structures, solitary waves can exist in the form of {\em quasiperiodic 
nonlinear localised modes}. As an example
we consider second-harmonic generation (SHG) and nonlinear beam 
propagation in {\em Fibonacci optical superlattices}, and demonstrate 
numerically the possibility of spatial self-trapping of quasiperiodic 
waves whose envelope amplitude varies quasiperiodically, while still 
maintaining a stable, well-defined spatially localised structure, 
{\em a quasiperiodic envelope soliton}.

We consider the interaction of a fundamental wave (FW) with the 
frequency $\omega$ and its second harmonic (SH) in a slab 
waveguide with quadratic (or $\chi^{(2)}$) nonlinearity. 
Assuming the $\chi^{(2)}$ susceptibility to be modulated and the 
nonlinearity to be of the same order as diffraction, we write 
the dynamical equations in the form
\begin{equation}
\label{dynam}
\begin{array}{l} 
  {\displaystyle i\frac{\partial w}{\partial z} + \frac{1}{2} 
  \frac{\partial^2 w}{\partial x^2} + d(z)w^*v e^{-i\beta z} 
  =  0,}\\*[9pt] 
  {\displaystyle i\frac{\partial v}{\partial z} + \frac{1}{4} 
  \frac{\partial^2 v}{\partial x^2} + d(z) w^2 e^{i\beta z} =  0,}
\end{array}
\end{equation}
where $w(x,z)$ and $v(x,z)$ are the slowly varying envelopes of the
FW and SH, respectively.
The parameter $\beta= \Delta k |k_{\omega}|x_0^2$ is proportional to 
the phase mismatch $\Delta k= 2k_{\omega}-k_{2\omega}$, $k_{\omega}$ 
and $k_{2\omega}$ being the wave numbers at the two frequencies. 
The transverse coordinate $x$ is measured in units of the input beam 
width $x_0$, and the propagation distance $z$ in units of the 
diffraction length $l_d= x_0^2|k_{\omega}|$. 
The spatial modulation of the $\chi^{(2)}$ susceptibility 
is described by the quasi-phase-matching (QPM) grating function 
$d(z)$. 
In the context of SHG, the QPM technique is an effective 
way to achieve phase matching, and has been studied intensively 
(see Ref.~\cite{QPM} for a comprehensive review).

Here we consider a QPM grating produced by a quasiperiodic nonlinear
optical superlattice. 
Quasiperiodic optical superlattices, one-dimensional analogs of 
quasicrystals \cite{quasi}, are usually designed to study the effect
of Anderson localisation in the linear regime of light propagation. 
For example, Gellermann {\em et al.} measured the optical transmission 
properties of quasiperiodic dielectric multilayer stacks of SiO$_2$ 
and TiO$_2$ thin films and observed a strong suppression of the
transmission \cite{geller}. 
For QPM gratings, a nonlinear quasiperiodic superlattice of
LiTaO$_3$, in which two antiparallel ferro-electric domains are 
arranged in a Fibonacci sequence, was recently fabricated by Zhu 
{\em et al.} \cite{Fib_exp}, who measured multi-colour SHG with 
energy conversion efficiencies of $\sim 5\%-20\%$.
This quasiperiodic optical superlattice in LiTaO$_3$ can 
also be used for efficient direct third harmonic generation
\cite{THG}.

\begin{figure}
\hspace{-10mm}\hbox{\epsfig{file=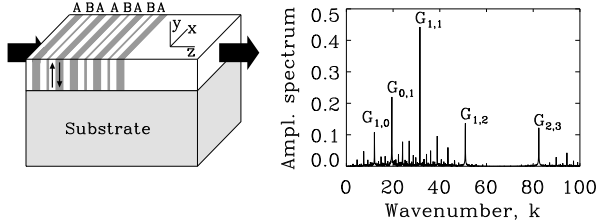,width=9.0cm}}
\hspace{3mm}
\caption{
(a) Slab waveguide with quasiperiodic QPM superlattice structure 
composed of building blocks A and B.
(b) Numerically calculated amplitude spectrum of $d(z)$.}
\label{d(z)}
\end{figure}

The quasiperiodic QPM gratings have two building blocks A and B 
of the length $l_A$ and $l_B$, respectively, which are ordered 
in a Fibonacci sequence [Fig.~\ref{d(z)}(a)]. 
Each block has a domain of length $l_{A_1}$=l ($l_{B_1}$=l) with
$d$=$+1$ (shaded) and a domain of length $l_{A_2}$=$l(1+\eta)$ 
[$l_{B_2}$=$l(1-\tau\eta)$] with $d$=$-1$ (white).
In the case of $\chi^{(2)}$ nonlinear QPM superlattices this 
corresponds to positive and negative ferro-electric domains, 
respectively.
The specific details of this type of Fibbonacci optical superlattices 
can be found elsewhere (see, e.g., Ref.~\cite{Fib_exp} and references
therein).
For our simulations presented below we have chosen $\eta$= $2(\tau-1)
/(1+\tau^2)$= 0.34, where $\tau$= $(1+\sqrt{5})/2$ is the so-called 
{\em golden ratio}. 
This means that the ratio of length scales is also the golden ratio, 
$l_A/l_B$= $\tau$. 
Furthermore, we have chosen $l$=0.1.

The grating function $d(z)$, which varies between $+1$ and $-1$
according to the Fibonacci sequence, can be expanded in a Fourier 
series
\begin{equation}
   \hspace{-7mm}
   d(z)= \sum_{m,n}d_{m,n}e^{iG_{m,n}z},\;\;\;
   G_{m,n}= \frac{2\pi(m+n\tau)}{D},
   \label{Gmn}
\end{equation}
where $D$=$\tau l_A+l_B$=0.52 for the chosen parameter values.
Hence the spectrum is composed of sums and differences of
the basic wavenumbers $\kappa_1$=$2\pi/D$ and $\kappa_2$=$2\pi\tau/D$.
These components fill the whole Fourier space densely, since $\kappa_1$
and $\kappa_2$ are incommensurate. Figure \ref{d(z)}(b) shows the
numerically calculated Fourier spectrum $G_{m,n}$. 
The lowest-order ``Fibonacci modes'' are clearly the most intense. 
From Eq.~(\ref{Gmn}) and the numerically found spectrum we identify 
the six most intense modes presented in Table 1.
The corresponding wavenumbers $G_{m,n}$ are in good agreement 
with Eq.~(\ref{Gmn}).

\begin{center}
\begin{tabular}{|l|llllll|}\hline
   $m$       &1     &0     &1     &2     &1     &2  \\
   $n$       &1     &1     &2     &3     &0     &4  \\
   $G_{m,n}$ &31.42 &19.42 &50.83 &82.25 &12.00 &101.66\\\hline
\end{tabular}
\end{center}

{\small TABLE 1. The six most intense Fibonacci modes $G_{m,n}$.}

To analyse the beam propagation and SHG in a quasiperiodic QPM 
grating one could simply average Eqs.~(\ref{dynam}).
To lowest order this approach always yields a system of
equations with constant mean-value coefficients, which does not 
allow to describe oscillations of the beam amplitude and phase. 
However, here we wish to go beyond the averaged equations and consider
the rapid large-amplitude variations of the envelope functions. 
This can be done analytically for periodic QPM gratings \cite{qpm}.
However, for the quasiperiodic gratings we have to resolve 
to numerical simulations.

\begin{figure}
\mbox{\epsfig{file=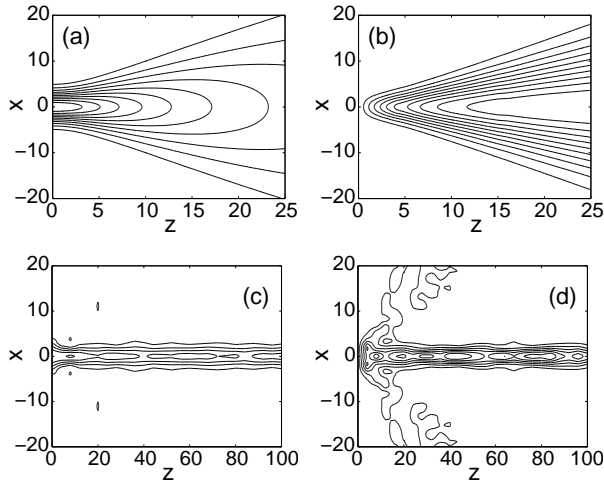,width=8cm}}
\caption{(a) Diffraction of a weak FW beam with amplitude $A_w$=0.25 
for $\beta$=82.25. (c) Excitation of a quasiperiodic soliton by a FW 
beam with amplitude $A_w$=5 for $\beta$=82.25. (b,d) Corresponding 
SH components.}
\label{soliton}
\end{figure}

Thus we have solved Eqs.~(\ref{dynam}) numerically with a second-order
split-step routine, in which the linear part is solved with the
fast-Fourier-transform (FFT) method and the nonlinear part, with a
fourth-order Runge-Kutta scheme. 
The step-length is adapting to the local domain length of the QPM 
grating. 
At the input of the crystal we excite the fundamental beam
(corresponding to unseeded SHG) with a Gaussian profile, 
\begin{equation}
   w(x,0) =  A_w \, e^{-x^2/10}, \;\;\; v(x,0) =  0.
   \label{initial}
\end{equation}
We consider the quasiperiodic QPM grating with matching to the 
peak at $G_{2,3}$, i.e., $\beta$=$G_{2,3}$=82.25. 
First, we study the small-amplitude limit when a weak FW is injected 
with a low amplitude. 
Figures \ref{soliton}(a,b) show an example of the evolution of FW 
and SH in this effectively linear regime.
As is clearly seem from Fig. \ref{soliton}(b) the SH wave is excited, 
but both beams eventually diffract.

When the amplitude of the input beam exceeds a certain threshold,
self-focusing and localization should be observed for both harmonics.
Figures \ref{soliton}(c,d) show an example of the evolution of a strong
input FW beam, and its corresponding SH. 
Again the SH is generated, but now the nonlinearity is so
strong that it leads to self-focusing and mutual self-trapping of the 
two fields, resulting in a spatially localized two-component soliton,
despite the continuous scattering of the quasiperiodic QPM 
grating.

\begin{figure}
\mbox{\epsfig{file=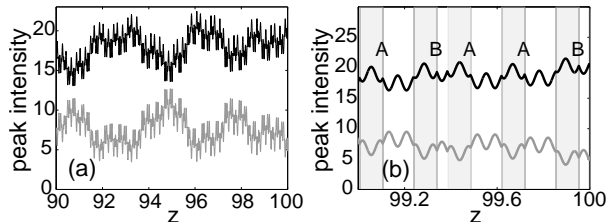,width=8cm}}
\caption{Amplitude oscillations of the quasiperiodic soliton.
(a),(b) Close-ups of the peak intensity $|w(0,z)|^2$ of the FW 
(black) and $|v(0,z)|^2$ of the SH (grey). 
The Fibonacci building blocks A and B are indicated in (b) with 
$d$=1 in grey regions, and $d$=$-1$ in white regions.}
\label{oscillations}
\end{figure}

It is important to notice that the two-component localised beam 
created due to the self-trapping effect is quasiperiodic by itself. 
As a matter of fact, after an initial transient its amplitude
oscillates in phase with the quasiperiodic QPM modulation $d(z)$. 
This is illustrated in Fig.~\ref{oscillations}, where we show 
in more detail the peak intensities in the
asymptotic regime of the evolution.

\begin{figure}
\mbox{\epsfig{file=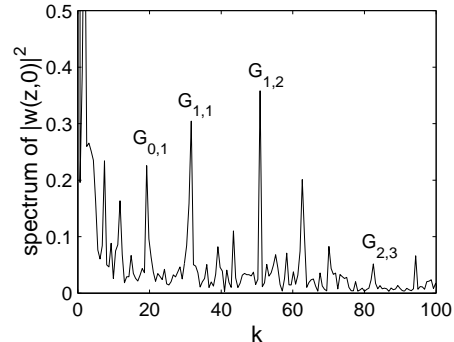,width=6cm}}
\caption{Spectrum of the amplitude oscillations of the FW component 
of the quasiperiodic soliton, calculated from $z$=90 to 100 in
Fig.~\ref{oscillations}(a). The peaks correspond to the Fibonacci peaks
$G_{m,n}$ in $d(z)$ and sum and difference thereof with the mismatch
$\beta$=82.25.}
\label{spectrum}
\end{figure}

Since the oscillations shown in Fig.~\ref{oscillations} are in phase
with the oscillations of the QPM grating $d(z)$, their spectra should 
be similar. This is confirmed by Fig.~\ref{spectrum}, which gives the 
spectrum of the peak intensity $|w(z,0)|^2$ of the FW.
Note that the Fibonacci peak at $k$=82.25 is suppressed (or reduced)
because the identical mismatch $\beta$ down-converts it to the 
dc-component.
Sum and difference wavenumbers between $\beta$ and $G_{m,n}$ appear,
which are generated by the nonlinearity. 
For example, the component at $k$=62.8 is the difference between 
$\beta$=82.25 and $G_{0,1}$=19.42.

Our numerical results show that the quasiperiodic envelope solitons 
can be generated for a broad range of the phase-mismatch $\beta$.
The amplitude and width of the solitons depend on the effective
mismatch, which is the separation between $\beta$ and the nearest
strong peak $G_{m,n}$ in the Fibbonacci QPM grating spectrum.
Thus, low-amplitude broad solitons are excited for $\beta$-values
in between peaks, whereas high-amplitude narrow solitons are
excited when $\beta$ is close to a strong peak, as shown in
Fig.~\ref{soliton}(c,d).

The existence of spatially localized self-trapped states in nonlinear
quasiperiodic media should not depend on the particular kind of
nonlinearity. 
The dependence on $\beta$ observed here for the $\chi^{(2)}$
gratings is simply because the "real" strength of the $\chi^{(2)}$
nonlinearity is inversely proportional to the phase-mismatch.
In fact, it is well-known that for large values of the mismatch 
$\beta$ the quadratic nonlinearity becomes effectively cubic 
\cite{kiv}.
Thus, our findings are directly applicable to nonlinear optical 
superlattices in cubic (or $\chi^{(3)}$) nonlinear media.

\begin{figure}
\mbox{\epsfig{file=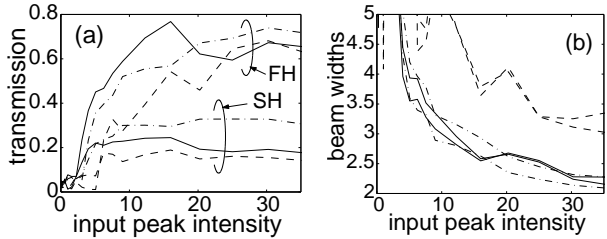,width=8.0cm}}
\caption{
(a) Transmission of the FW, $|w(0,L)/w(0,0)|^2$, and the SH,
$|v(0,L)/w(0,0)|^2$ vs.\ input peak intensity $|w(0,0)|^2$ 
of the FW. (b) Output beam width vs.\ $|w(0,0)|^2$. $L$=100, 
$\beta$=$G_{1,2}$=50.83 (solid), $\beta$=$G_{2,4}$=101.66
(dashed), and $\beta$=$G_{2,3}$=82.25 (dot-dashed).}
\label{transmission}
\end{figure}

To analyse in more detail the transition between the linear 
(diffraction) and nonlinear (self-trapping) regimes, we have made
a series of careful numerical simulations. 
In Fig.~\ref{transmission} we show the transmission coefficients and 
the beam widths at the output of the crystal versus the intensity 
of the FW input beam, for a variety of $\beta$-values.
These dependencies clearly illustrates the 
universality of the generation of localised modes for varying strength 
of nonlinearity, i.e. a quasiperiodic
soliton is generated only for sufficiently high amplitudes. 
This is of course a general phenomenon also observed in many nonlinear 
isotropic media. 
However, here the self-trapping occurs for quasiperiodic waves, with
the quasiperiodicity being preserved in the variation of the amplitude 
of both components of the soliton.

Numerical simulations for other values of the phase-mismatch $\beta$ 
reveal the same basic property of quasiperiodic self-trapping: 
Spatial solitons are formed in Fibonacci quadratic nonlinear slab 
waveguides above a certain power threshold, and such solitons are 
always {\it quasiperiodic}, i.e. they exhibit large-amplitude 
oscillations along $z$, which are composed of mixing of the two 
incommensurate Fibonacci wavenumbers $\kappa_1$ and $\kappa_2$. 
The amplitude and width of these solitons depend on the difference 
between the phase-mismatch $\beta$ and the nearest strong peak 
$G_{m,n}$ in the Fibonacci spectrum. 

Finally we would like to emphasize that the phenomenon described here 
is qualitatively different from the propagation of topological and
nontopological {\em kinks} in disordered and quasiperiodic nonlinear 
media \cite{gredeskul}. 
Such kinks can be well approximated by an effective structureless 
particle, which either preserves identity, as in the case of 
topological kinks \cite{gred,sanchez}, or decays rapidly into 
radiation \cite{swed}.

In conclusion, we have analysed SHG, self-focusing, and nonlinear 
beam propagation in Fibonacci optical superlattices with a quadratic 
nonlinear response. 
We have predicted spatial self-trapping of quasiperiodic waves and 
the formation of quasiperiodic solitons. 
Such solitons have a localised envelope that traps the random-phase 
components through the phase and frequency locking effect of strong 
nonlinearity, and whoes amplitude undergoes clearly detectable
quasiperiodic oscillations. 
The results presented here would allow to extend the concepts of 
self-localisation and self-modulation of nonlinear waves to a broader 
class of spatially inhomogeneous media, and can also be found in 
systems of different physical context.

The authors acknowledge support from 
the Danish Technical Research Council (Talent Grant no.~9800400), 
the Danish Natural Science Research Council (Grant no. 9600852), and 
the Department of Industry, Science, and Tourism (Australia).

\end{multicols}


\begin{references}
\bibitem{book}
   See, e.g., {\em ``Future Directions of Nonlinear Dynamics in 
   Physical and Biological Systems''}, P.L. Christiansen {\em et 
   al.}, Eds., NATO ASI Series B: Physics, Vol. 312 (Plenum Press,
   New York, 1993).

\bibitem{hasegawa}
   A. Hasegawa, Phys. Fluids {\bf 18}, 77 (1975); Phys.
   Fluids {\bf 20}, 2155 (1977).

\bibitem{inc_exp}
   M. Mitchell {\em et al.}, Phys. Rev. Lett. {\bf 77}, 490 (1996); 
   see also M. Mitchell and M. Segev, Nature {\bf 387}, 880 (1997); 
   Z. Chen {\em et al.}, Science {\bf 280}, 889 (1998).

\bibitem{gredeskul}
   See, e.g., S.A. Gredeskul and Yu.S. Kivshar,  
   Phys. Rep. {\bf 216}, 1 (1992), and references therein.

\bibitem{my_prl}
   Yu.S. Kivshar, S.A. Gredeskul, A. S\'anchez, and L. V\'azquez,
   Phys. Rev. Lett. {\bf 64}, 1693 (1990).

\bibitem{hopkins}
   V.A. Hopkins, J. Keat, G.D. Meegan, T. Zhang, and J.D. Maynard,
   Phys. Rev. Lett. {\bf 76}, 1102 (1996).

\bibitem{QPM} 
   M.M. Fejer, G.A. Magel, D.H. Jundt, and R.L. Byer,
   IEEE J. Quantum Electron. {\bf 28}, 2631 (1992).

\bibitem{quasi} 
   D. Schechtman, I. Blech, D. Gratias, and J.W. Cahn,
   Phys. Rev. Lett. {\bf 53}, 1951 (1984).

\bibitem{geller} 
   W. Gellermann, M. Kohmoto, B. Sutherland, and P.C. Taylor,
   Phys. Rev. Lett. {\bf 72}, 63 (1994).

\bibitem{Fib_exp}
   S. Zhu, Y. Zhu, Y. Qin, H. Wang, C. Ge, and N. Ming,
   Phys. Rev. Lett. {\bf 78}, 26752 (1997).

\bibitem{THG}
   S. Zhu, Y. Zhu, and N. Ming, Science {\bf 278}, 843 (1997).

\bibitem{qpm} 
   C. Balslev Clausen, O. Bang, and Yu.S. Kivshar,
   Phys. Rev. Lett. {\bf 78}, 4749 (1997).

\bibitem{kiv} 
   See, e.g., Yu.S. Kivshar, In: {\em Adavanced Photonics with
   Second-order Optically  Nonlinear Processes}, Eds. A.D. Boardman
   {\em et al.} (Kluwer, Amsterdam, 1999), p. 451.

\bibitem{gred} 
   S.A. Gredeskul {\em et al.}, Phys. Rev. A {\bf 45}, 8867 (1992).

\bibitem{sanchez} 
   F. Dominguez-Adame, A. S\'anchez, and Yu.S. Kivshar,
   Phys. Rev. E {\bf 52}, 1283 (1995).

\bibitem{swed} 
   See, e.g., M. H\"ornquist and R. Riklund, 
   J. Phys. Soc. Jpn. {\bf 65}, 2872 (1996).
\end{references}
\end{document}